\documentclass[preprint2]{aastex}

\shorttitle{Disk M/L determination }
\shortauthors{Repetto et al.}

\begin{document}

\title{Disk Mass-to-Light ratio distribution 
from stellar population synthesis: Application to rotation curve 
decomposition of NGC~5278 (KPG~390 A).}

\author{P. Repetto\altaffilmark{1}, Eric E. Mart\'inez-Garc\'ia\altaffilmark{1,2},
M. Rosado\altaffilmark{1}, R. F. Gabbasov\altaffilmark{1}}

\email{prsatch6@gmail.com}

\altaffiltext{1}{Instituto de Astronom\'{\i}a, Universidad Nacional 
Aut\'onoma de M\'exico (UNAM),\\ Apdo. Postal 70-264, 04510, 
M\'exico, D.F., M\'exico.}
\altaffiltext{2}{Instituto Nacional de Astrof\'isica, 
\'Optica y Electr\'onica (INAOE), Aptdo. Postal 51 y 216, 72000 Puebla
, Pue., M\'exico.}

\begin{abstract}
In this work we extend the study on the mass distribution of the spiral galaxy NGC 5278,
performing 1D and 2D (GALFIT) bulge-disk decomposition to determine which components
constitute the baryonic mass in this galaxy. Our analysis does not detect any bulge, instead 
we find a bright source, probably related with the central AGN, and an 
exponential disk. We fix the stellar disk contribution to the rotation 
curve (RC) with broad band photometric observations and population synthesis 
models, to obtain the 2D mass distribution of the stellar disk. In particular, for NGC 5278, 
we find that the typical assumption of considering the mass-to-luminosity ratio ($M/L$) of the disk 
as constant along the galactocentric radius is not valid. We also extract a baryonic RC from the mass 
profile, to determine the inability of this baryonic RC (also taking into account $\pm$ 30$\%$ errors 
in the disk mass), to fit the entire RC. We perform the RC decomposition of NGC 5278 considering the 
determined baryonic RC and four types of dark matter (DM) halo: Hernquist, Burkert, Navarro, 
Frenk, \& White and Einasto. Our results determine that Hernquist DM halo better models our 
observed RC in the case of disk mass $M_d=5.6\times10^{10}$ $M_{\odot}$ and also with less 30$\%$ 
disk mass. In the case of more 30$\%$ disk mass the cored Einasto ($n < 4$) DM halo is
the best fitting model.
\end{abstract}


\section{Introduction.}\label{sec1}

NGC 5278 is a spiral galaxy participating in an interaction process with the companion NGC 5279. Both galaxies constitute
the isolated galaxy pair KPG 390 that belongs to the Karachentsev catalog of isolated pairs of galaxies in the northern hemisphere 
(KPG) \citep{Karachentsev1972} and it is classified as a M 51 galaxy type by \citet{Klimanov2001}. The pairs of galaxies in the KPG 
catalog are in the first stage of interactions and in particular KPG 390 being considered M 51 galaxy type is not in a merger status. 
NGC 5278 and NGC 5279 are Markarian galaxies and \citet{Keel1985} have classified the two galaxies as intermediate type between
Seyferts and liners, this means that the center of NGC 5278 is the host of an Active Galactic Nucleus (AGN). Another important 
information we obtain from the literature is that the morphological classification of NGC 5278 is not unique: Sb 
(Hyperleda\footnote{http://leda.univ-lyon1.fr} \citep{Paturel2003}), S0 D (Simbad) and SA(s)b? pec (NED).\\
We have already performed a detailed study of the interaction process in a previous work \citep{Repetto2010}. In our earlier study 
we derive velocity fields and RCs of both members of KPG 390. The hypothesis to build the RC from our H$\alpha$
velocity fields is that in the relation of the line of sight velocity the terms of radial velocity and vertical velocity are
small enough with respect to the terms of rotational velocity. In the case of NGC 5278 we have already encountered that condition 
inside an angular sector of $20^{\circ}$ around the kinematic major axis of this galaxy (the residuals are less than 60 km $s^{-1}$ 
within that angular sector of $20^{\circ}$) and we derive the RC of NGC 5278 within this angular sector to remove the effect of 
non-circular motions. Inside that sector, the RC of NGC 5278 really traces the gravitational potential required to decompose 
its RC and the corresponding parameters ($V_{sys}$, P.A. and inclination) are free from any uncertainty derived from possible 
regions of non-circular motions in NGC 5278.\\ 
In general the RCs of spiral galaxies are important tools used to infer the 
contribution to the gravitational potential of the different components of a galaxy: stellar disk, bulge and bar, dust, gas, and 
DM halo \citep{vanderKruit1989, Sofue2001}. As a matter of fact, the existence of very massive DM halos 
was inferred from the flatness of the RCs at large galactocentric distances \citep{Rubin1980}. The kinematic data used to derive the 
RCs could come from absorption lines of the stars in the disk of the galaxy, from emission lines of the ionized gas in HII 
regions of the disk, or from neutral gas in the disk.\\ 
The method to study the DM halo distribution takes the RC, constructed 
from the kinematic observations of a spiral galaxy, and expresses it as $V^2(R)$, where $V^2(R)$ is the 
square of the rotational velocity derived from the kinematics and $R$ is the galactocentric distance. Then it fits the 
contribution to $V^2(R)$ of the assumed different components such as: a stellar (exponential) thin disk, a stellar bulge, 
the HI gas, and a DM halo with different possible radial density profiles. A non linear least squares minimization is used to select 
which disk and halo density distributions fit better the observed kinematics. Several important assumptions are made in using this 
method. Let us discuss the ones concerning the stellar disk contribution to the kinematics. One of the major problems arises because 
from the photometry we obtain only the surface brightness distribution, not the surface mass density distribution. This implies that 
we have to assume a certain $M/L$ ratio of the stellar disk to convert surface brightness to mass density. In our earlier work we 
accomplished the RC decomposition of NGC 5278, considering a disk component without performing bulge-disk decomposition, and setting 
free the $M/L$ ratio of the disk (within acceptable values), because we do not have any constraint of that value. In that former 
study we performed the RC decomposition considering three DM halos: Hernquist halo \citep{Hernquist1990a}, Pseudo-Isothermal Halo 
(P-ISO) \citep{vanAlbada1985} and Navarro, Frenk White Halo (NFW) \citep{Navarro1996}. In \citet{Repetto2010} we were unable to 
discriminate between the three DM halos employed to decompose the RC of NGC 5278. The three halos reproduce the RC of NGC 5278 well 
enough with different disk masses and scale lengths and the values obtained for the halo masses and radii are in accordance with the 
typical values reported by cosmological numerical simulations \citep{Navarro1996}.\\ 
In all the previous studies on galaxy mass determination through RC decomposition, the $M/L$ ratio of the disk was considered 
constant along the radius and taken as a free parameter. One of the first approaches was proposed by \citet{vanAlbada1985} who 
varied the disk $M/L$ from its minimum value (zero, i.e., no stars in the disk) to its maximum value (the one that fitted the inner
RC). This approach is known as ``the maximum-disk fit''. For instance, \citet{Blais-Ouellette1999} study the Sd galaxy NGC 5585, 
considering the M/L ratio of the stellar disk as a free parameter, in the process of mass decomposition of the RC of this galaxy. 
The authors obtained that the P-ISO DM halo succeeded in reproducing the observed RC of NGC 5585, nonetheless the resulting $M/L$ 
ratio of the stellar disk that better fits the different components had unrealistic low values ($M/L \sim 0.3-0.8$). 
Since then, much work \citep{deBlok2001, Swaters2003, Spano2008, Chemin2011} has been done on the mass distribution of galaxies 
through RC decomposition, always considering the M/L ratio as a free parameter, leaving the principal recipe ($M/L \sim $const) 
unchanged; nevertheless, using this simple recipe, the M/L ratio values are often unphysical. 
\citet{Fuentes-Carrera2007} have tried to fix the $M/L$ of the disk with more reasonable values obtained from the disk color indexes, 
comparing them with those of stars and adopting the $M/L$ of those stars as stellar disk $M/L$. The authors obtained as best fit DM 
halo the P-ISO DM profile, however they consider that the disk is composed of stars of only one spectral type; instead a 
detailed stellar population synthesis study is necessary, to better constrain the disk $M/L$.\\ 
The motivation of the present study is to extend the RC decomposition of NGC 5278, this time with new constraints 
on the baryonic mass of this galaxy, derived from additional observations concerning surface photometry in several bands and 
using stellar population synthesis studies \citep{Zibetti2009} to determine the baryonic $M/L$ of NGC 5278.  
From the broad band surface photometry analysis and the stellar population synthesis we obtain the disk mass profile of NGC 5278. 
From that mass profile we derive the baryonic RC of NGC 5278 by means of the cumulative integration of the surface 
brightness along the line of sight, to establish the inadequacy of this baryonic RC and also of the baryonic RC more or less 30$\%$ 
disk mass to account for the entire mass in NGC 5278. In the mass modeling of NGC 5278 we also explore a disk mass variation of more
or less 30$\%$ considering four types of DM halos: Hernquist, Burkert \citep{Burkert1995}, NFW and Einasto 
\citep{Einasto1965}. In this analysis we find that the RC of NGC 5278 can be reproduced well enough by Hernquist DM halo with a disk 
mass $M_d=5.6\times10^{10}$ $M_{\odot}$ and also in the case of less 30$\%$ disk mass. The cored Einasto ($n < 4$) DM halo produced 
the best fit to the RC of NGC 5278 with more 30$\%$ disk mass.\\
The article is divided as follows: In the introduction we present the problem of fitting several components to galaxy RCs derived 
from kinematical observations. In section~\ref{sec2} we accomplish bulge-disk decomposition to establish the baryonic mass component 
of NGC 5278, showing that there is no bulge contribution. Section~\ref{sec3} addresses the method to obtain disk mass 
distributions from certain color indexes and population synthesis models. This section also gives the application of this method to 
obtain the stellar disk mass distribution of NGC 5278. In Section~\ref{sec4} we describe the procedure to derive the baryonic RC of 
NGC 5278 from the mass profile obtained in Section~\ref{sec3}. In Section~\ref{sec5}, once we have determined the true disk mass 
distribution, we use the baryonic disk of NGC 5278 to fit only the DM halo distribution in the observed RC. In Section~\ref{sec6} we
consider the maximum disk solution for NGC 5278, where we utilize the baryonic disk determined in section~\ref{sec4} to fit the 
majority of internal points of the observed RC. In Section~\ref{sec7} a discussion of our results and their implications is presented 
and the conclusions are outlined in section~\ref{sec8}.

\section{1D and 2D bulge-disk decomposition of NGC 5278.}\label{sec2}

\begin{figure*}[htp!]
\centering
\epsscale{2.1}
\plotone{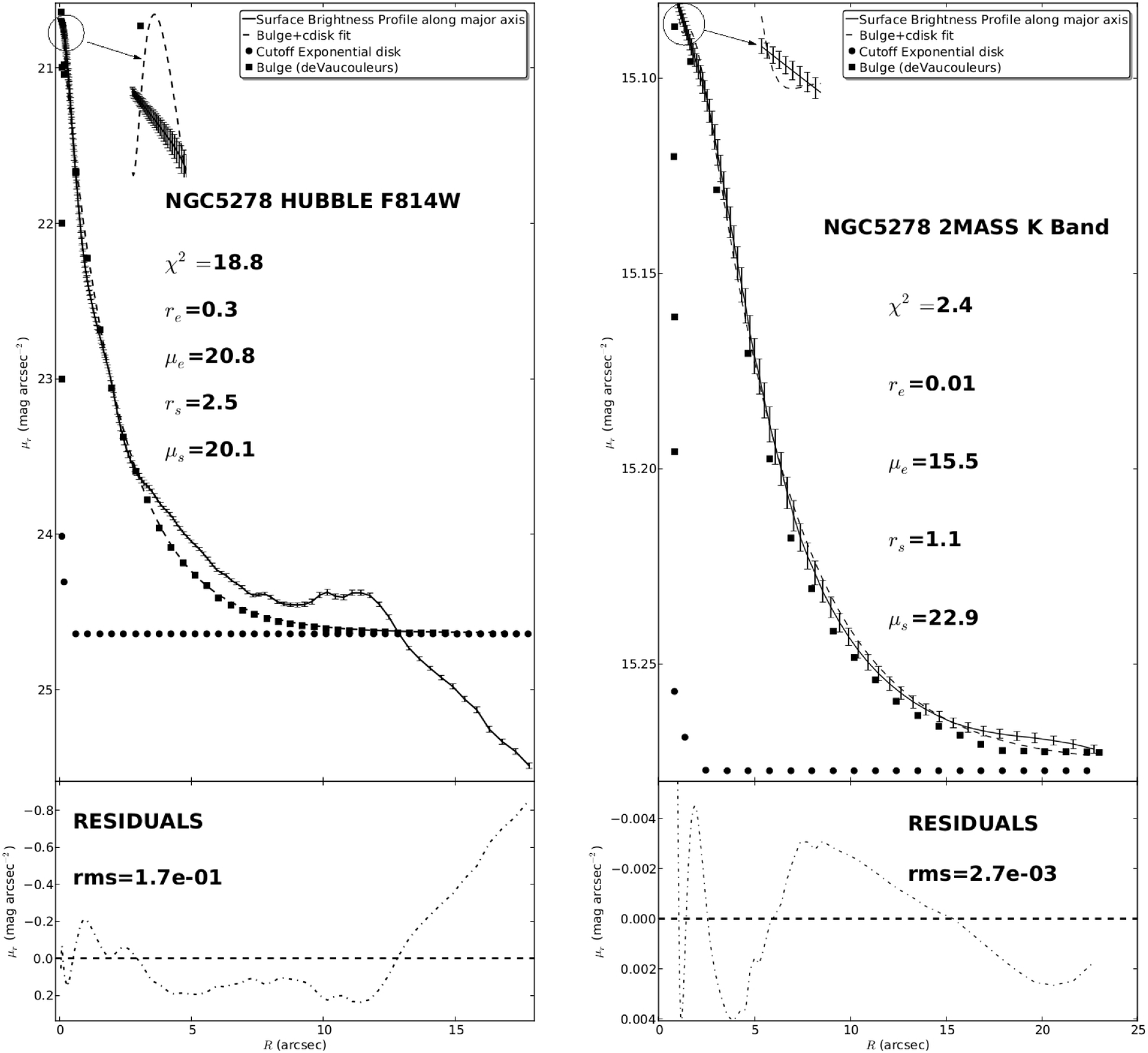}
\caption{Surface brightness profiles fit along major axis for the Hubble F814W and 2MASS K band. The ``cdisk'' nomenclature in the 
figures is referred to the cutoff exponential disk expressed by the relation $I(r)=I_0\exp{[-(\alpha r +\beta^3 r^{-3})]}$, 
where $\beta$ has length dimensions and $I(r)$ reduces to an exponential disk for $r \gg \beta$. This cutoff disk profile, 
according to \citet{Kormendy1977}, is necessary to better fit the underlying disk component. The zoom in the upper part of the 
figures shows how the deVaucoleurs fit to the expected bulge of NGC 5278 does not succeed.}
\label{fig1}
\end{figure*}

As we already mentioned in the Introduction the morphological classification of NGC 5278 is uncertain and in the Hyperleda database
it is classified as Sb, so we want to explore the possibility of the presence of a bulge component in NGC 5278. In the Introduction
we have also reminded that NGC 5278 shows an AGN in its center, fact that could create more confusion in the morphological 
classification. For these reasons we perform 1D bulge-disk decomposition of NGC 5278 considering several photometric bands 
in three ground-based telescopes (DSS (B), SDSS (u, g, i), 2MASS (J, K)) and in the HST (F300W, F814W). For all 
telescopes and photometric bands considered we obtain the same result, in this section we only show the surface brightness profile 
fit to the 2MASS K band and Hubble F814W. We made use of IRAF\footnote{IRAF is distributed by the National Optical Astronomy 
Observatory, which is operated by the Association of Universities for Research in Astronomy (AURA) under cooperative agreement with 
the National Science Foundation.} task {\it ellipse} to obtain the surface brightness profiles along the major and minor axes of 
NGC 5278, and of IRAF task {\it nfit1d} to fit a deVaucouleurs \citep{deVaucouleurs1948} bulge and an exponential disk 
\citep{Freeman1970} according to the procedure devised by \citet{Kormendy1977}. We make a PyRAF-based script to run both IRAF tasks 
for all the photometric bands considered. In this section we only show the surface brightness profile fit along the major axis 
because the profiles along major and minor axes are identical but the profile along major axis is more extended. The results are 
displayed in the Figure~\ref{fig1} and it is clear from the two panels that the deVaucouleurs profile do not fit the inner part of 
the surface brightness profile i.e., the plausible bulge component of this galaxy.

\begin{figure*}[htp!]
\centering
\epsscale{2.1}
\plotone{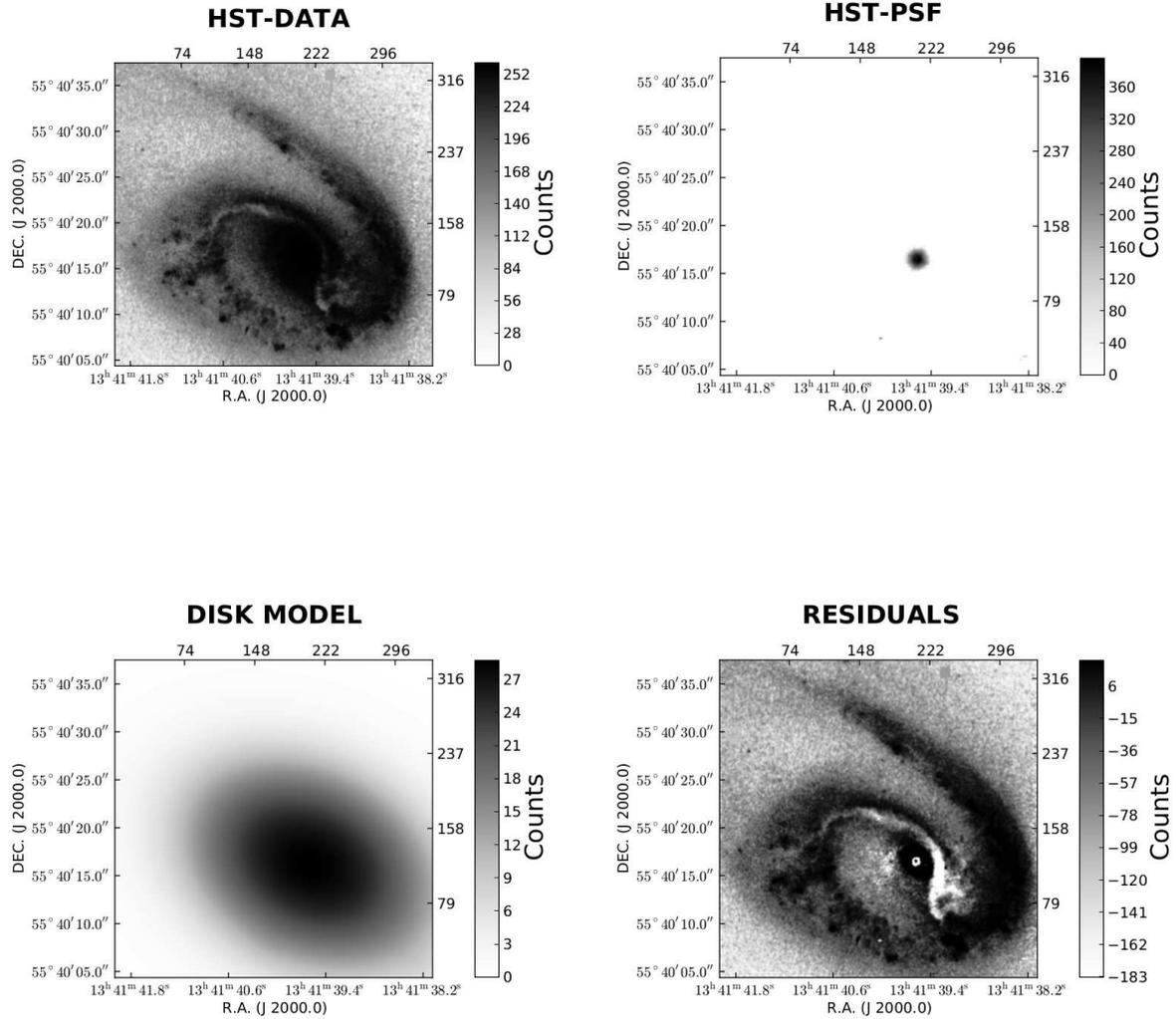}
\caption{GALFIT 2D decomposition of the NGC 5278 F814W Hubble image. In the residual image 
the spiral arms mean level is about 18 counts with a maximum of 23 counts, in this figure 
there is clearly visible the dust lane, more evident that in the original data where the disk 
emission of NGC 5278 partially hides this feature. In the same image the counts level around the 
central PSF is about 20.}
\label{fig2}
\end{figure*}

\citet{Kormendy2004} demonstrated that Sb and earlier-type galaxies contain principally classical bulges (deVaucouleurs), whereas Sbc 
galaxies have more often non-classical bulges (Sersic index between 1 and 2) and Sc and later type galaxies seem never to have 
classical bulges. These facts are very important and given the incongruity in the morphological classification of NGC 5278 we are 
obliged to use a Sersic profile \citep{Sersic1968} to fit the expected bulge component in NGC 5278. For this purpose we perform 2D 
bulge-disk decomposition using GALFIT\footnote{http://users.obs.carnegiescience.edu/peng/work/galfit/galfit.html} 
\citep{Peng2002, Peng2010}. The principal reason to utilize GALFIT is because that software 
makes a 2D decomposition of a galaxy image allowing to employ several components such as exponential disk, Sersic, deVaucouleurs, 
Gaussian, King profile, Moffat, PSF and other components for bars, spiral arms and Fourier modes. 
The presence of a central AGN, i.e., a point source, and the fact that GALFIT cannot resolve sources with radius less than 0.5 pixels 
(corresponding to 0.025 kpc for NGC 5278) and axis ratio less than 0.1, forced us to perform the fit with the F814W Hubble image of 
NGC 5278 due to the high resolution (0.0996$\arcsec$ $\sim$ 0.005 kpc for NGC 5278) of Hubble images and the fact that GALFIT needs 
high resolution images to work appropriately. The motivation to choose the F814W filter, which is equivalent to the Johnson I band, 
is substantially due to the lack of NICMOS infrared images for NGC 5278 that could better demonstrate the existence of a classical 
bulge in NGC 5278. In fact, in the case of this galaxy there are only images from the WFPC2 in the three filters F814W, F255W and 
F300W. NGC 5278 is practically invisible in the filter F255W that is in the mid-UV, in the filter F300W there is considerable 
emission of this galaxy but F300W is in the near-UV, so the more suitable choice for our purpose (considering a classical bulge made 
up of old stars) is the filter F814W. We begin to fit a PSF function (fwhm 5.5 pixels $\approx$ 0.27 kpc) to eliminate the strong 
central compact peak and we succeeded to remove most of the central source (fwhm 7.4 pixels $\approx$ 0.37 kpc) proceeding then to 
fit an exponential disk component to the residuals of the first fit (85 counts). After removing other relics of the AGN
we manage to get residuals of almost 23 counts, similar to the counts of the spiral arms. In fact, the brighter spot of this last 
residual map is represented by an HII region in the periphery of the spiral arms (23 counts). In these last residuals, the zone 
around the extracted central point source has 20 counts. Thus, the bulge-disk decomposition analysis in NGC 5278 do not detect any 
bulge component, instead we find a bright, almost punctual, central spot in the center, probably associated to the central AGN. The 
results of our GALFIT analysis are shown in Figure~\ref{fig2}. 

\clearpage

\section{Determining the disk M/L ratio of NGC 5278 from stellar
population synthesis models.}\label{sec3}

In order to determine the mass profile and total mass estimate for NGC 5278 in an appropriate way,
resolved stellar mass maps are needed. Structural parameters may change between luminosity and 
stellar mass studies because the mass to luminosity ratio, $M/L = (M_{*}/M_{\sun})/(L_{*}/L_{\sun})$, 
is not a constant throughout the disk. The method of \citet{Zibetti2009} 
is capable of reconstructing resolved stellar mass maps of galaxies
from multi-band optical/NIR imaging. The technique is based on a Monte Carlo library
of 50000 stellar population synthesis (SPS) models supported in the 2007 version of \citet{Bruzual2003} 
code. These models include a new prescription by \citet{Marigo2007} 
for the thermally pulsating asymptotic
giant branch (TP-AGB) evolution of intermediate and low mass stars. 
The library covers a wide range of star formation histories, metallicities, and dust contents.
The prescription to treat dust attenuation is
based on the two-component dust model of \citet{Charlot2000}.
In this prescription, the emission from
stars younger than $10^7$ yr, which is the time needed for the dissipation
of dense molecular clouds, is more attenuated than that from older stars.

\begin{figure*}[htp!]
\centering
\epsscale{2.1}
\plotone{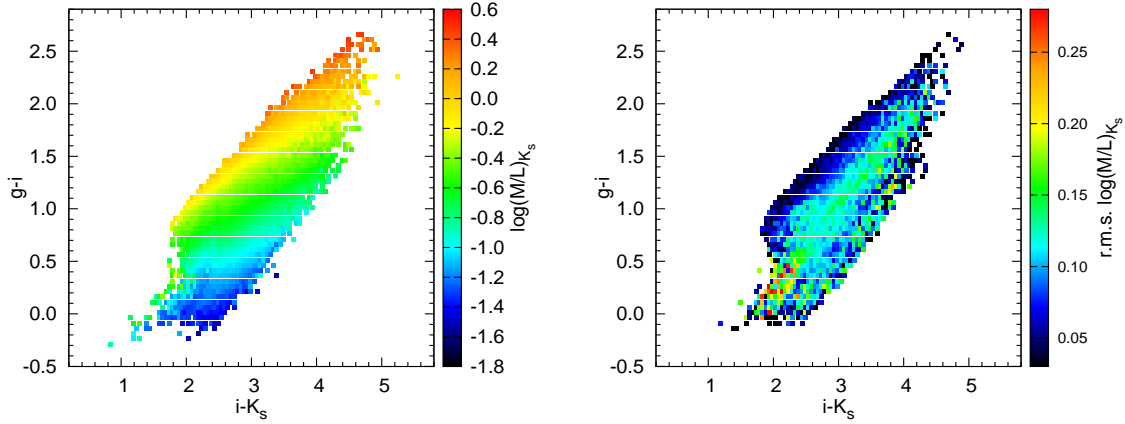}
\caption{Effective $M/L$ in $K_{s}$-band as a function of ($i-K_{s}$) and ($g-i$) colors at a redshift of $z = 0.025$.
Left panel: the median logarithmic $M/L$ for models binned $0.05 \times 0.05$ mag$^2$ in the color-color space. 
Right panel: the rms of log $M/L$.}
\label{fig3}
\end{figure*}

We take advantage of the novel technique of \citet{Zibetti2009},
and construct a spatially resolved map of stellar mass surface density for NGC 5278.
For this purpose we use the $g$-band and $i$-band data from SDSS DR8 \citep{Aihara2011},
and the $K_{s}$-band data from 2MASS \citep{Skrutskie1997,Skrutskie2006}.
The procedure is as follows. The Monte Carlo spectral library was obtained from \citet{daCunha2008}.
The MAGPHYS (Multi-wavelength Analysis of Galaxy Physical Properties) 
code\footnote{http://www.iap.fr/magphys/magphys/MAGPHYS.html} \citep{daCunha2008} was used to 
obtain lookup tables of ($i-K_{s}$) color,
($g-i$) color, and $M/L$ ratio at a redshift of $z = 0.025$ (see figure~\ref{fig3}).
Models were binned in 0.05 magnitude bins, and the median value for each bin was computed.
We compare the ($i-K_{s}$) and ($g-i$) colors obtained from the photometry with the values
obtained from the binned models on a pixel-by-pixel basis.
The resolved mass map was obtained by multiplying the $K_{s}$ luminosity image by the
$M/L$ ratio map.

For all calculations we adopt the cosmological parameters
$H_{0}=70.0$ km sec$^{-1}$ Mpc$^{-1}$, $\Omega_{\mathrm{matter}}=0.3$, and $\Omega_{\mathrm{vacuum}}$ 
or $\Omega_{\lambda}=0.7$. The luminosity distance adopted from NED was $(m-M) = 35.24$ mag, or 112 Mpc.
The absolute magnitude of the Sun in the $K_{s}$-band was taken from \citet{Blanton2007}.
No corrections for Galactic extinction were made since $E(B-V)$ = 0.009 mag \citep{Schlegel1998}.

\begin{figure*}[htp!]
\centering
\epsscale{2.1}
\plotone{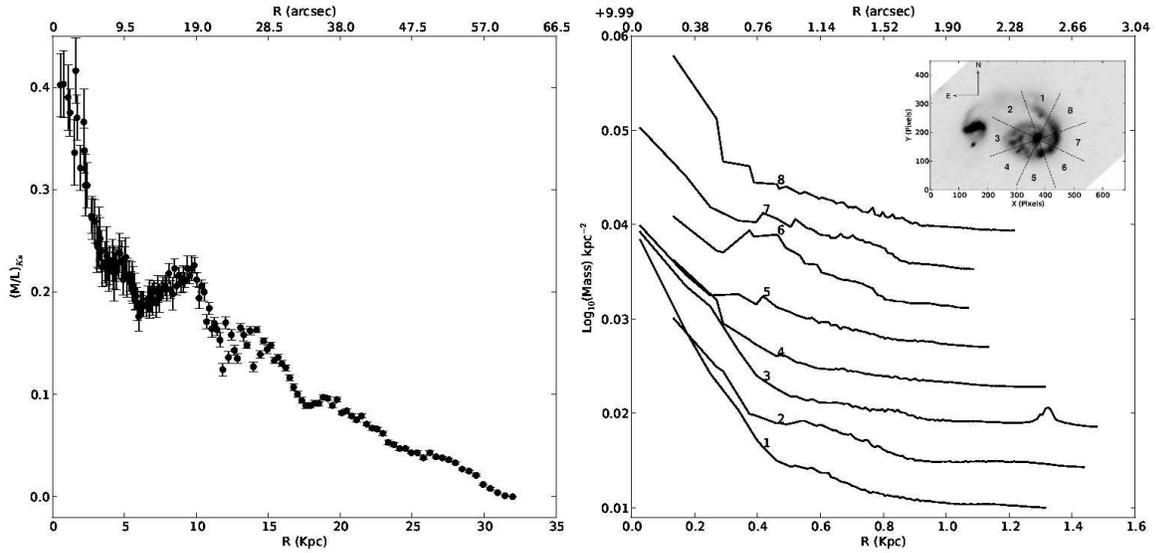}
\caption{
{\it Left}: Azimuthally averaged radial profile of $M/L$ in 
$K_{s}$-band for NGC 5278, adopting~\citet{Zibetti2009} method based on $g$, $i$, and 
$K_{s}$ photometry. {\it Right}: Radial mass profiles of NGC 5278 plotted in logarithmic 
scale, with the KPG 390 image superimposed to denote the regions where the profiles 
are taken in the galaxy disk of NGC 5278. The radial profiles were shifted vertically 
by the logarithmic amount of 10$^{9}$.}
\label{fig4}
\end{figure*}

The resolved stellar mass map was deprojected adopting the projection parameters
of \citet{Repetto2010}. The resulting mass to light (in the $K_{s}$-band) ratio and the mass 
(azimuthally averaged) profiles are shown in figure~\ref{fig4}. 

\begin{figure*}[htp!]
\centering
\epsscale{2.1}
\plotone{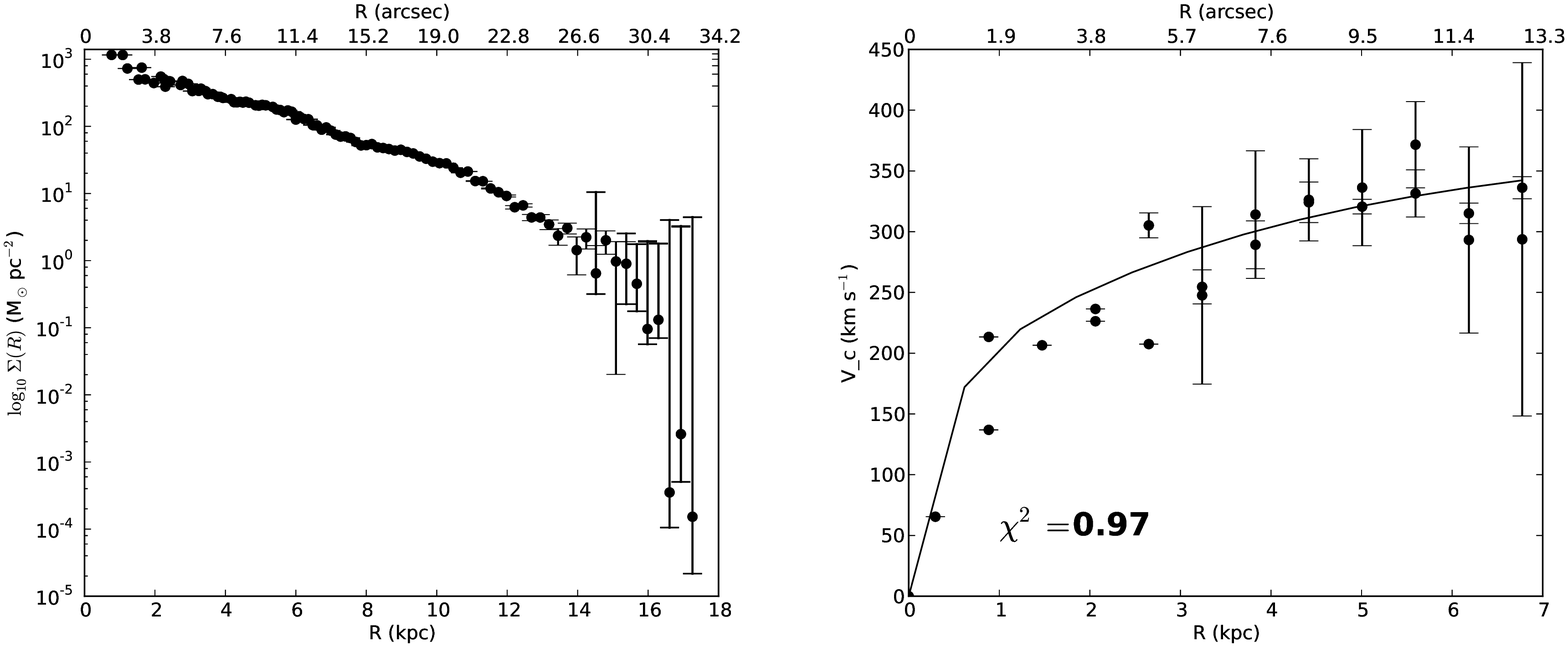}
\caption{{\it Left}: Mean logarithmic mass profile extracted using the method outlined 
in section~\ref{sec3}. {\it Right}: H$\alpha$ observed rotation curve of NGC 5278, obtained
in \citet{Repetto2010}, preliminary fitted with an exponential function.\label{fig5}}
\end{figure*}

\section{The necessity of DM to decompose the RC of NGC 5278.}\label{sec4}

\begin{figure*}[!htp]
\epsscale{1.0}
\plotone{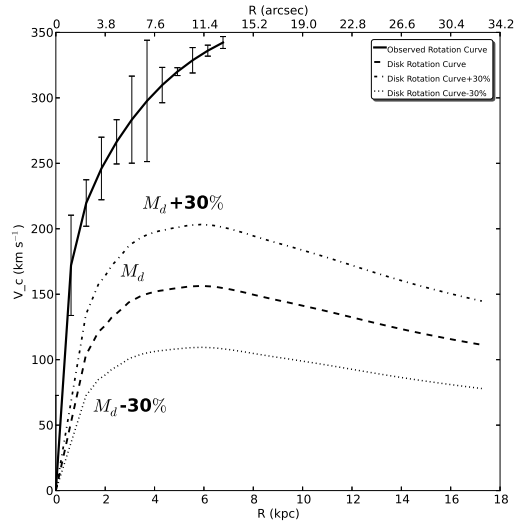}
\caption{Baryonic disk of NGC 5278 more 
and less 30$\%$ disk mass. In this panel is also shown the total RC (disk+DM) of 
NGC 5278 for comparison. 
\label{fig6}}
\end{figure*}

From the mass profile of figure~\ref{fig5} (left panel) we derive the baryonic part of the disk of NGC 5278, 
considering the relation $M(R)=2\pi\int_0^r \Sigma(R) R dR$ to compute the surface mass density inside the radial extensions of 
the mass profile of figure~\ref{fig5} (33 arcsecs $\approx$ 17.3 kpc), then we calculate the circular 
velocity with the relation $V^2(R)=GM(R)/R$. We also analyze a variation of the baryonic part of the disk of NGC 5278 of more or 
less 30$\%$ disk mass, due to a 30$\%$ intrinsic uncertainties\footnote{The biggest uncertainty comes from propagating the error 
in the assumed distance to the object.} of the \citet{Zibetti2009} method. The result is shown in 
figure~\ref{fig6}. From this figure we observe that if we only consider the baryonic disk of NGC 5278 there is needed at least 70$\%$ 
more mass to recover the RC of this galaxy. The fact that the whole RC of NGC 5278 requires the DM component to explain the mass 
distribution of this galaxy is not obvious from the fit we performed in our previous work, because in that earlier study we randomly 
varied the $M/L$ ratio of the baryonic disk of NGC 5278. In that former study we fitted jointly the free disk $M/L$ component with 
the DM component and we obtained different values for the disk mass and scale radius for the different DM halos (P-ISO, Hernquist 
halo \citep{Hernquist1990a} and NFW) we have analyzed \citep{Repetto2010}. 

\noindent That procedure determines the true parameters of the 
baryonic component of NGC 5278, nevertheless, the actual baryonic mass of NGC 5278 could have been overestimated or underestimated, 
as well as the DM component of that galaxy. In the present analysis we avoid this problem constraining the disk parameters by means 
of broad band surface photometry additional observations and stellar population synthesis studies so that we only set free the DM 
parameters.

\section{RC decomposition of NGC 5278 fitting only the DM component 
after subtraction of the determined baryonic disk.}\label{sec5}

\begin{figure*}[!htp]
\epsscale{2.1}
\plotone{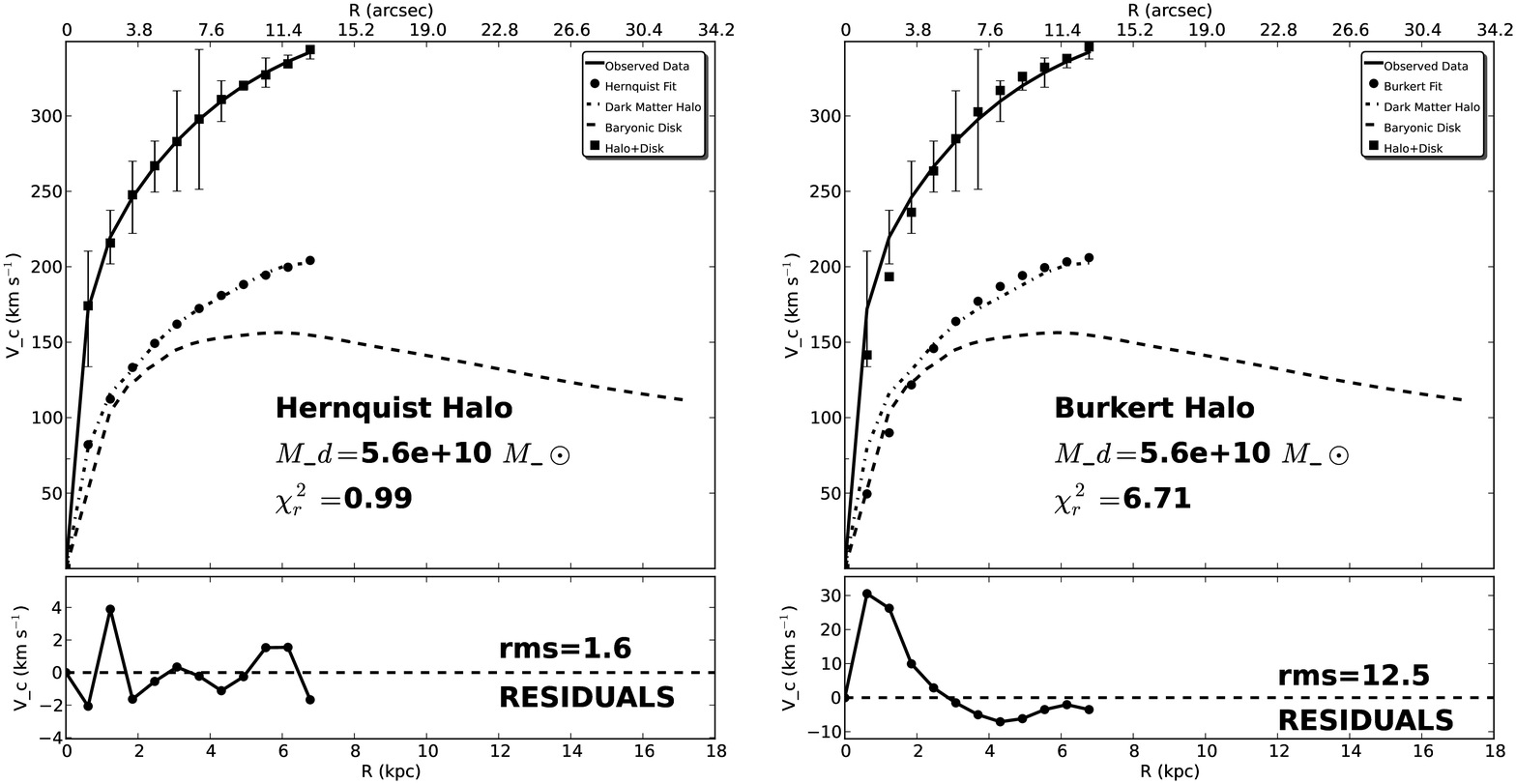}
\caption{Rotation curve decomposition of NGC 5278 with disk 
mass $M_d=5.6\times10^{10}$.\label{fig7}}
\end{figure*}

\begin{figure*}[!htp]
\epsscale{2.1}
\plotone{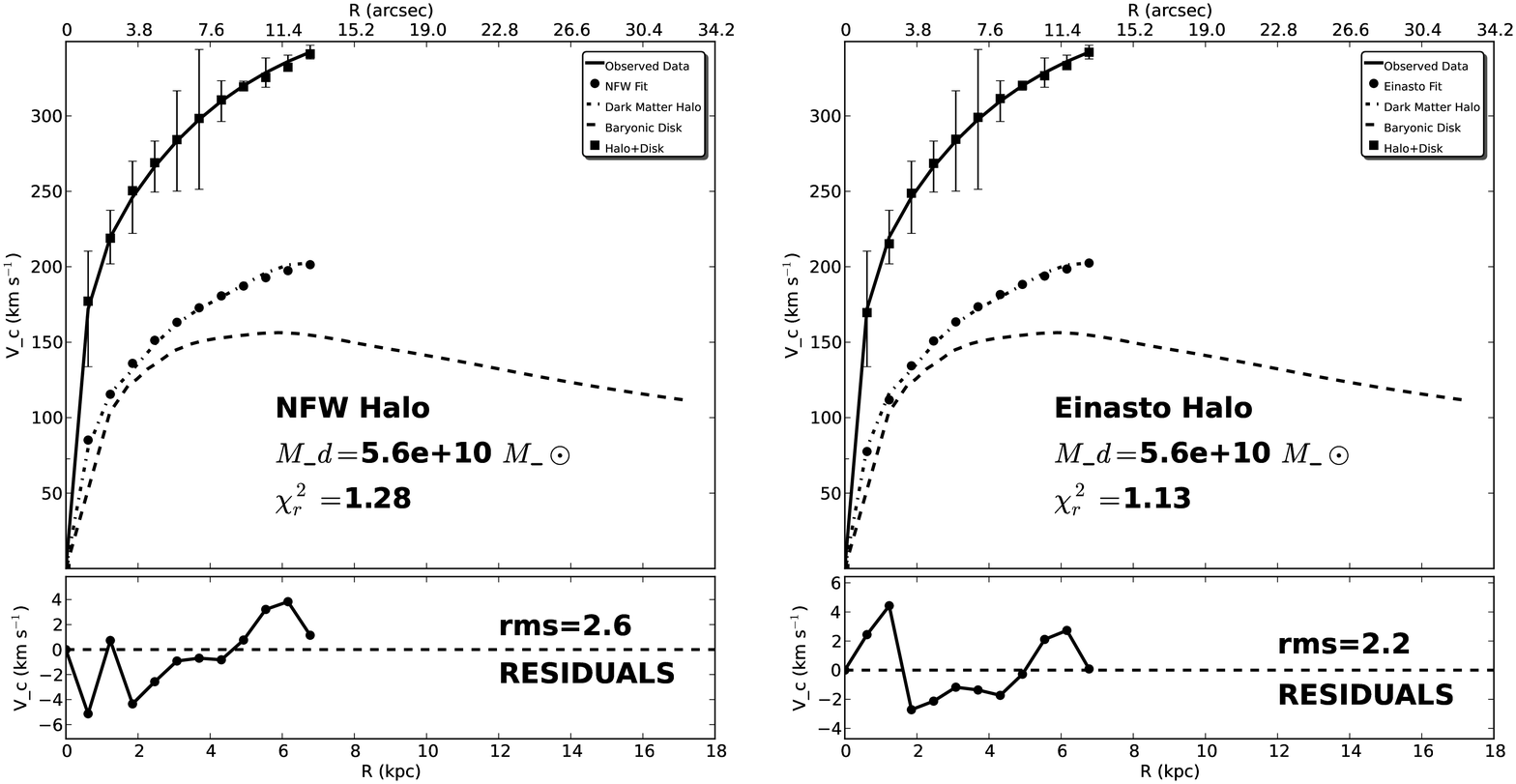}
\caption{Rotation curve decomposition of NGC 5278 with disk 
mass $M_d=5.6\times10^{10}$.\label{fig8}}
\end{figure*}

\begin{figure*}[!htp]
\epsscale{2.1}
\plotone{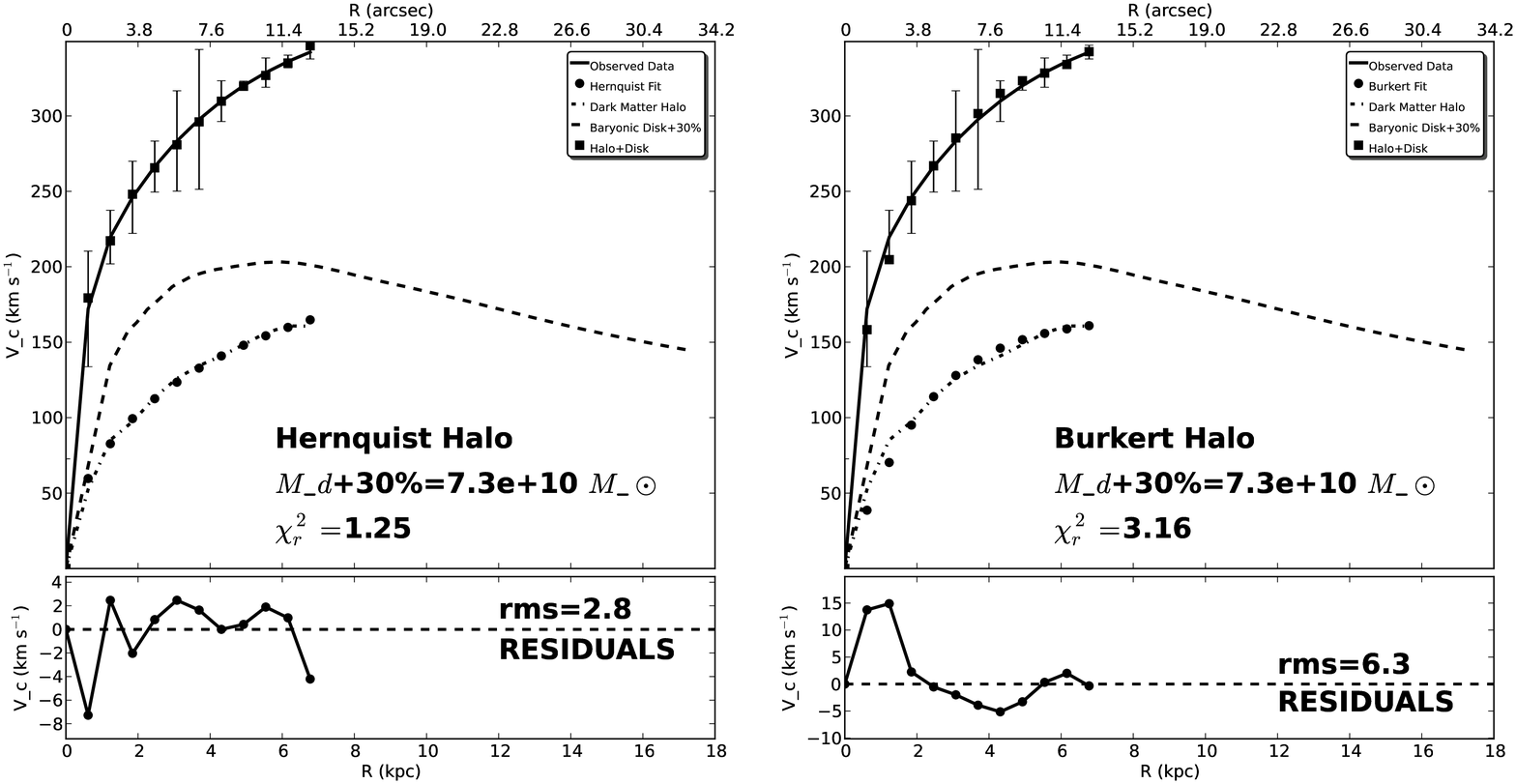}
\caption{Rotation curve decomposition of NGC 5278 with 
more 30$\%$ disk mass $M_d=7.3\times10^{10}$.\label{fig9}}
\end{figure*}

\begin{figure*}[!htp]
\epsscale{2.1}
\plotone{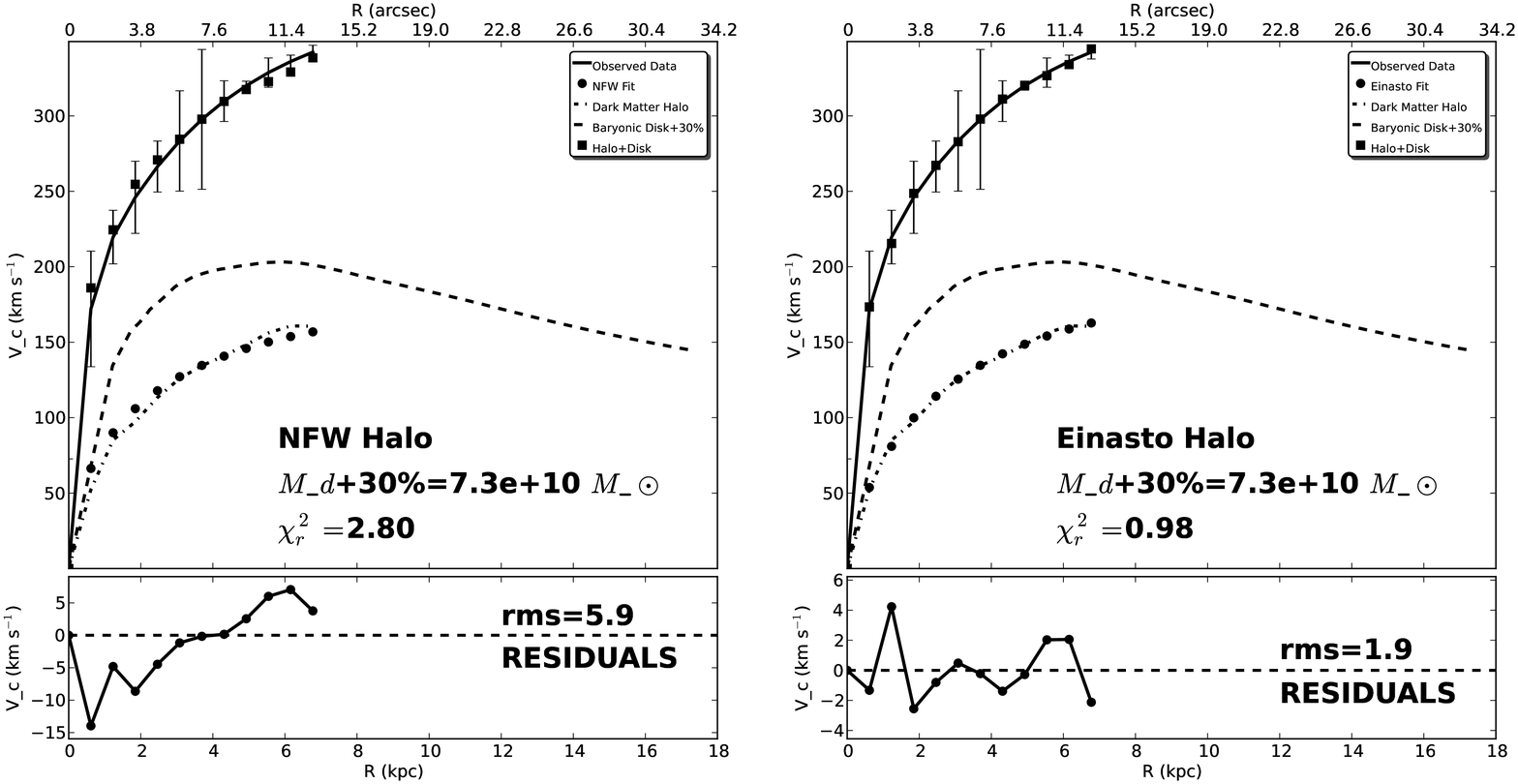}
\caption{Rotation curve decomposition of NGC 5278 with 
more 30$\%$ disk mass $M_d=7.3\times10^{10}$.\label{fig10}}
\end{figure*}

\begin{figure*}[!htp]
\epsscale{2.1}
\plotone{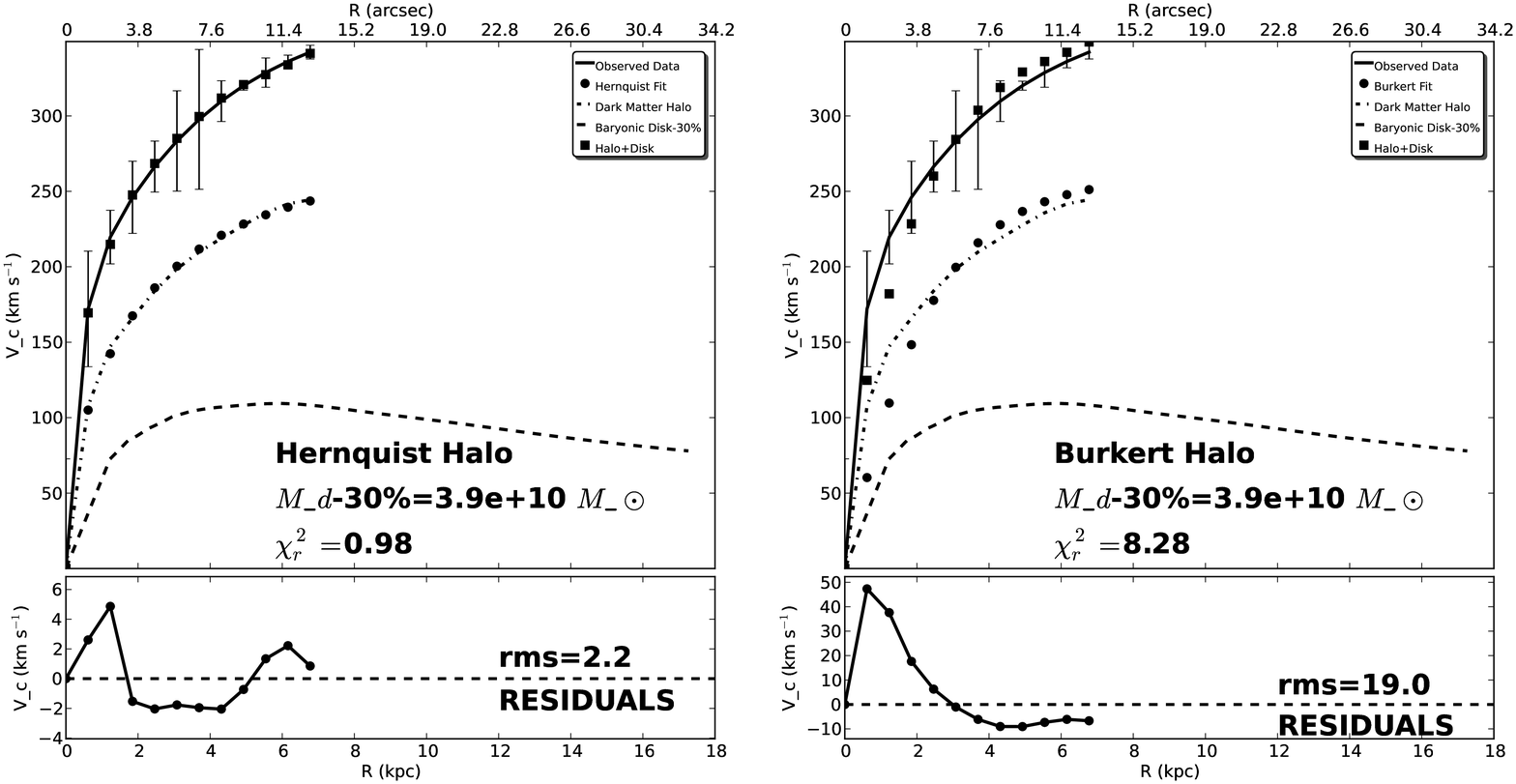}
\caption{Rotation curve decomposition of NGC 5278 with 
less 30$\%$ disk mass $M_d=3.9\times10^{10}$.\label{fig11}}
\end{figure*}

\begin{figure*}[!htp]
\epsscale{2.1}
\plotone{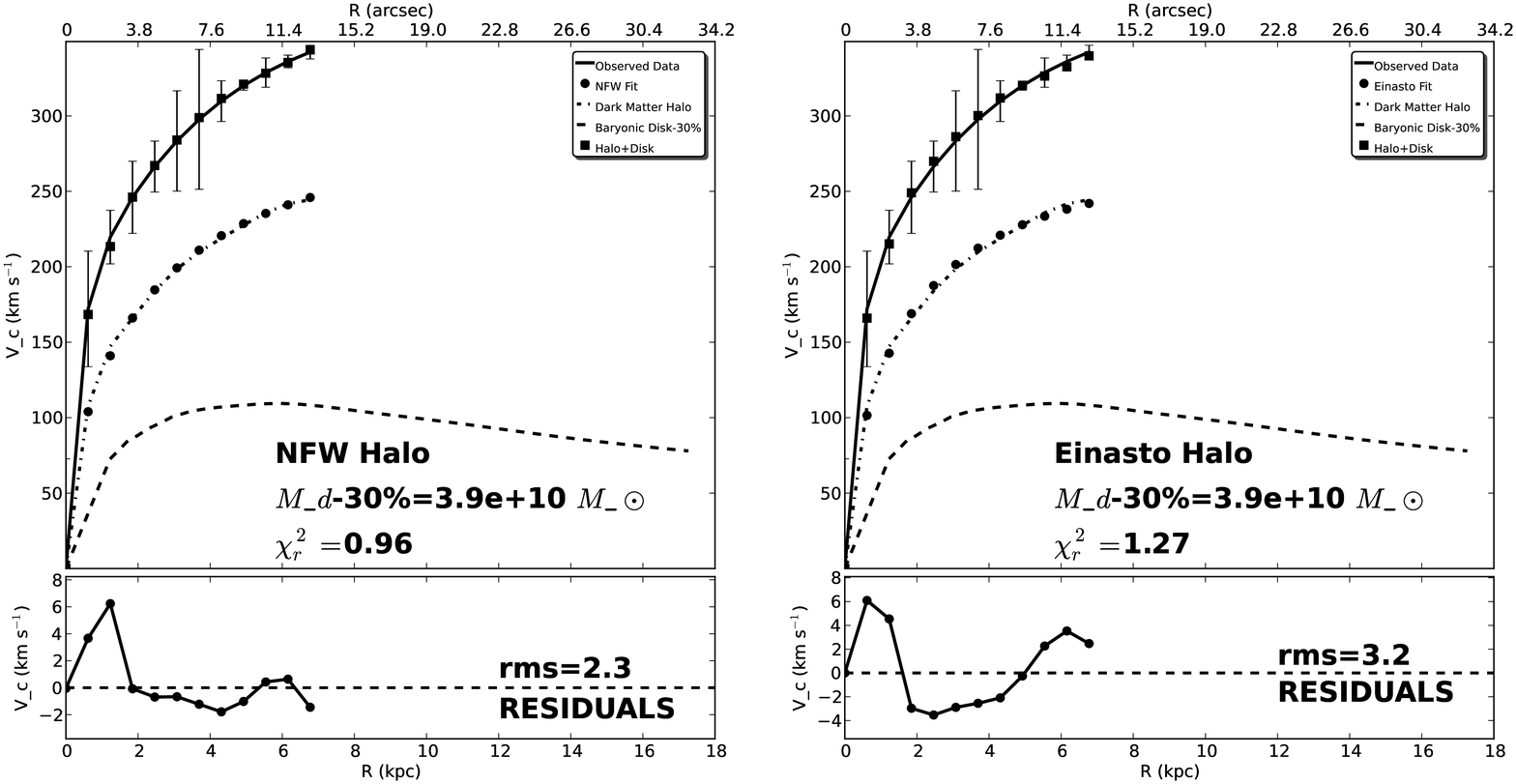}
\caption{Rotation curve decomposition of NGC 5278 with 
less 30$\%$ disk mass $M_d=3.9\times10^{10}$.\label{fig12}}
\end{figure*}

The RC decomposition was performed  using the module mpfit\footnote{We use the python Kmpfit version from the Kapteyn package 
\citep{KapteynPackage} that uses the C implementation of mpfit \citet{Markwardt2009}. The corresponding website is 
http://purl.com/net/mpfit.} (Levemberg-Marquardt method). We extract the DM RC of NGC 5278 subtracting the baryonic RC of that
galaxy displayed in figure~\ref{fig6} from the observed RC of NGC 5278, also taking into account a disk mass variation of more 
and less 30$\%$ to consider a 30$\%$ uncertainty in the determination of the disk mass. The
corresponding total baryonic mass estimates derived from the baryonic RC of NGC 5278 are respectively 5.6$\times$10$^{10}$ 
$M_{\odot}$ (disk mass), 7.3$\times$10$^{10}$ $M_{\odot}$ (disk mass more 30$\%$) and 3.9$\times$10$^{10}$ $M_{\odot}$ 
(disk mass less 30$\%$).\\ 
We proceed to explore the DM halo distribution by fitting the DM RC of NGC 5278. We consider this time four halo models: Hernquist, 
Burkert, NFW and Einasto \citep{Einasto1965}). In the present work we add two new halo profiles, the Burkert DM halo 
\citep{Burkert1995} and the Einasto DM halo with three free parameters \citep{Einasto1965}. The reason to utilize Burkert DM halo 
instead of P-ISO is dictated by the fact that Burkert DM halo is more general than P-ISO and behaves like P-ISO (i.e. it presents 
a core) at smaller radii while converges to a NFW at larger radii. The expression of the spatial density of the Burkert halo is the 
following \citep{Burkert1995}:

\begin{equation}\label{eq:0}
\rho_B(r)=\frac{\rho_0 r^3_0}{(r+r_0)(r^2+r^2_0)}, 
\end{equation}

\noindent where $\rho_0$ and $r_0$ are free parameters that represent the central DM density and the scale radius.
The choice to include the Einasto DM halo is motivated by relatively recent $\Lambda$-CDM dissipationless simulations of gravitational 
clustering and isolated spherical collapse \citep{Navarro2004, Navarro2010, Tissera2010}. This DM halo density profile has already 
been used by other authors to fit the DM distribution when baryons are included \citep{Onorbe2007, Gao2008}.
\citet{Navarro2004, Navarro2010} proposed a model with functional form $\rho\left(r\right) \propto \exp{\left(-Ar^{\alpha}\right)}$, 
where $\alpha$ is a parameter describing the degree of curvature of the profile. 
\citet{Merritt2005} pointed out that this is the same relation previously proposed by Einasto to describe the spatial 
density distribution of galaxies \citep{Einasto1965}. It is important to note that this DM density profile introduces more free 
parameters so that, in principle, it could allow a better fit to the RCs. For example, \citet{Chemin2011} fit 20 RCs of the THINGS 
survey of isolated galaxies, with the Einasto, NFW and P-ISO profile and the result of their analysis is that Einasto index of less 
than 4 (cored Einasto) produces a better fit with respect to P-ISO in about 60$\%$ of the RCs considered, while Einasto index 
greater than 4 (cuspy Einasto) produces a better fit with respect to NFW in about 80$\%$ of the RCs. For all these reasons we use 
in this work the Einasto DM halo as well. \citet{Merritt2006} provide the following formulation for the spatial density of the 
Einasto DM halo:

\begin{equation}\label{eq:1}
\rho_E(r)=\rho_e\exp{\{-d_n[(r/r_e)^{1/n}-1]\}},
\end{equation}

\noindent where $\rho_e$ is the density at the radius $r_e$ that defines a volume containing half of the total mass, the term 
$d_n$ is a function of $n$ defined according to the approximation $d_n=3n-1/3+0.0079/n$ for $n \gtrsim 0.5$.\\ 
During the fitting procedure we leave free the DM halo radius and the DM halo mass (and also the Einasto index). 
In order to accomplish the fitting process, we required to start with initial DM halo guesses $M_h$ and $a_h$, where $M_h$ is the 
DM halo mass and $a_h$ is the DM halo radius. We vary these halos parameters within the intervals $M_h \in$[10$^{9}$-10$^{14}$] 
$M_{\odot}$, $a_h \in$[0.1-40.0] kpc. For the Einasto DM density profile the range of variation of 
the halo shape parameter is $n \in$[0.6-10.0]. The choice of these intervals is twofold: the necessity to explore DM halo parameters 
domain as wide as possible and the obligation to avoid possible regions where the parameters values could lose their physical 
meaning. We use these ranges for the DM halo masses and the DM halo radii also in the cases of more or less 30$\%$ disk mass.
From the fitting process we obtain that the best fit to the DM RC is the Hernquist halo profile with a disk mass 
$M_d=5.6\times10^{10}$ $M_{\odot}$ and less 30$\%$ of this disk mass value. For more 30$\%$ disk mass the best fitting model to the
DM RC of NGC 5278 is the cored ($n < 4$) Einasto DM halo , nevertheless the value of the Einasto index, of our analysis, is outside 
of the interval ($5 \le n_{sim} \le 7$) predicted by the numerical simulations of \citet{Navarro2004, Navarro2010}. The fits to the 
RC for all disk masses and the four types of DM halos are shown in figures~\ref{fig7}, \ref{fig8}, \ref{fig9}, \ref{fig10}, 
\ref{fig11}, \ref{fig12}. The values of the halo masses and halo radii are displayed in tables~\ref{tbl-1}, \ref{tbl-2}, \ref{tbl-3}. 
We can see from the masses values in these tables that the different mass models, in our study, have similar halo masses within a 
factor 4, and the concentrations for the NFW halo and Einasto halo are in accordance with the last simulation results 
\citep{Klypin2011, Prada2012}. From the three figures it is clear that the Burkert DM halo is not able to reproduce our observed RC, 
for the disk masses considered, and the final halo radius of 3 kpc indicates that it is equivalent to P-ISO DM halo, because as 
mentioned above the Burkert DM halo has a flat core at smaller radii. 

\clearpage

\clearpage

\begin{table}
\begin{center}
\caption{Rotation curve decomposition with $M_d=$5.6$\times$10$^{10}$ $M_{\odot}$.\label{tbl-1}}
\begin{tabular}{lrrrrr}
\tableline\tableline
Principal Results & Hernquist & Burkert & NFW & Einasto\tablenotemark{b} &\\
\tableline\tableline
Halo Mass ($M_{\odot}$) & 8.7$\times$10$^{11}$ & 5.5$\times$10$^{11}$ & 9.9$\times$10$^{11}$  & 1.6$\times$10$^{12}$  &\\
Halo Radius (kpc)  & 17.9 & 3.0 & 9.0 & 18.6 &\\
$\chi^2_r$\tablenotemark{a} & 0.99 & 6.71 & 1.28  & 1.13  &\\
Dof\tablenotemark{a} & 10 & 10 & 10  & 9  &\\
\tableline\tableline
Additional Results  & Initial $R_{vir}$ (kpc) &  Final $R_{vir}$ (kpc) & Initial $c_{vir}$  & Final $c_{vir}$  &\\
\tableline\tableline
NFW Halo  & 150.0 &  167.8 & 10.2  & 9.9  &\\
Einasto Halo & 133.8 &  198.1 & 10.5  & 9.6  &\\
\tableline\tableline
\end{tabular}
\tablenotetext{a}{Reduced $\chi^2$ and degrees of freedom.}
\tablenotetext{b}{Final Einasto index n=3.6.}
\end{center}
\end{table}

\begin{table}
\begin{center}
\caption{Rotation curve decomposition with $M_d=$7.3$\times$10$^{10}$ $M_{\odot}$ (more 30$\%$ disk mass).\label{tbl-3}}
\begin{tabular}{lrrrrr}
\tableline\tableline
Principal Results & Hernquist & Burkert & NFW & Einasto\tablenotemark{b} &\\
\tableline\tableline
Halo Mass ($M_{\odot}$) & 1.3$\times$10$^{12}$ & 3.3$\times$10$^{11}$ & 5.4$\times$10$^{11}$  & 1.1$\times$10$^{12}$  &\\
Halo Radius (kpc)  & 30.1 & 3.0 & 9.0 & 18.4 &\\
$\chi^2_r$\tablenotemark{a} & 1.25 & 3.16 & 2.80  & 0.98  &\\
Dof\tablenotemark{a} & 10 & 10 & 10  & 9  &\\
\tableline\tableline
Additional Results  & Initial $R_{vir}$ (kpc) &  Final $R_{vir}$ (kpc) & Initial $c_{vir}$  & Final $c_{vir}$  &\\
\tableline\tableline
NFW Halo  & 150.0 &  137.7 & 10.2  & 10.4  &\\
Einasto Halo & 133.8 &  175.0 & 10.5  & 9.9  &\\
\tableline\tableline
\end{tabular}
\tablenotetext{a}{Reduced $\chi^2$ and degrees of freedom.}
\tablenotetext{b}{Final Einasto index n=3.1.}
\end{center}
\end{table}

\begin{table}
\begin{center}
\caption{Rotation curve decomposition with $M_d=$3.9$\times$10$^{10}$ $M_{\odot}$ (less 30$\%$ disk mass).\label{tbl-2}}
\begin{tabular}{lrrrrr}
\tableline\tableline
Principal Results & Hernquist & Burkert & NFW & Einasto\tablenotemark{b} &\\
\tableline\tableline
Halo Mass ($M_{\odot}$) & 8.5$\times$10$^{11}$ & 8.2$\times$10$^{11}$ & 1.6$\times$10$^{12}$  & 2.1$\times$10$^{12}$  &\\
Halo Radius (kpc)  & 13.7 & 3.0 & 9.0 & 16.8 &\\
$\chi^2_r$\tablenotemark{a} & 0.98 & 8.28 & 0.96  & 1.27  &\\
Dof\tablenotemark{a} & 10 & 10 & 10  & 9  &\\
\tableline\tableline
Additional Results  & Initial $R_{vir}$ (kpc) &  Final $R_{vir}$ (kpc) & Initial $c_{vir}$  & Final $c_{vir}$  &\\
\tableline\tableline
NFW Halo  & 150.0 &  196.0 & 10.2  & 9.6  &\\
Einasto Halo & 133.8 &  215.6 & 10.5  & 9.4  &\\
\tableline\tableline
\end{tabular}
\tablenotetext{a}{Reduced $\chi^2$ and degrees of freedom.}
\tablenotetext{b}{Final Einasto index n=3.7.}
\end{center}
\end{table}

\clearpage

\begin{table}[!hb]
\begin{center}
\caption{Maximum disk solution for NGC 5278 $M_d=$2.1$\times$10$^{11}$ $M_{\odot}$.\label{tbl-4}}
\rotatebox{90}{
\begin{tabular}{lrrrrr}
\tableline\tableline
Principal Results & Hernquist & Burkert & NFW & Einasto\tablenotemark{b} &\\
\tableline\tableline
Halo Mass ($M_{\odot}$) & 7.5$\times$10$^{11}$ & 2.4$\times$10$^{11}$ & 6.3$\times$10$^{11}$  & 9.8$\times$10$^{11}$  &\\
Halo Radius (kpc)  & 21.6 & 59.0 & 57.0 & 20.7 &\\
$\chi^2_r$\tablenotemark{a} & 1.80 & 0.63 & 2.01  & 0.90  &\\
Dof\tablenotemark{a} & 10 & 10 & 10  & 9  &\\
\tableline\tableline
Additional Results  & Initial $R_{vir}$ (kpc) &  Final $R_{vir}$ (kpc) & Initial $c_{vir}$  & Final $c_{vir}$  &\\
\tableline\tableline
NFW Halo  & 150.0 &  144.2 & 10.2  & 10.3  &\\
Einasto Halo & 133.8 &  36.5 & 10.5  & 14.0  &\\
\tableline\tableline
\end{tabular}}
\tablenotetext{a}{Reduced $\chi^2$ and degrees of freedom.}
\tablenotetext{b}{Final Einasto index n=1.8.}
\end{center}
\end{table}

\section{Maximum Disk solution for NGC 5278.}\label{sec6}

\begin{figure*}[!htp]
\epsscale{2.1}
\plotone{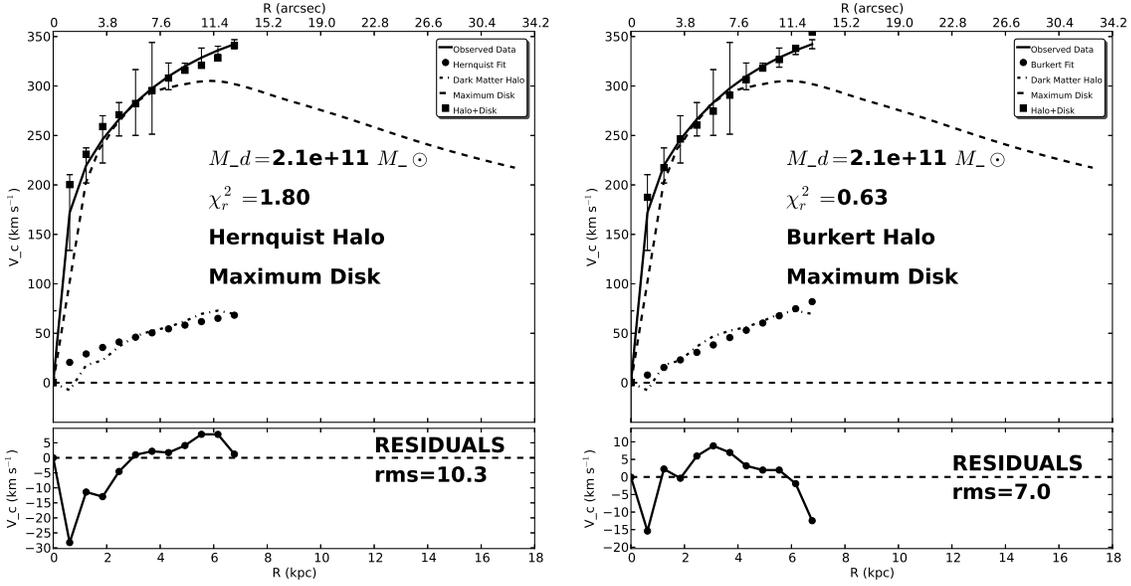}
\caption{Maximum disk solution with $M_d=$ 
2.1$\times$10$^{11}$ $M_{\odot}$.\label{fig13}}
\end{figure*}

\begin{figure*}[!htp]
\epsscale{2.1}
\plotone{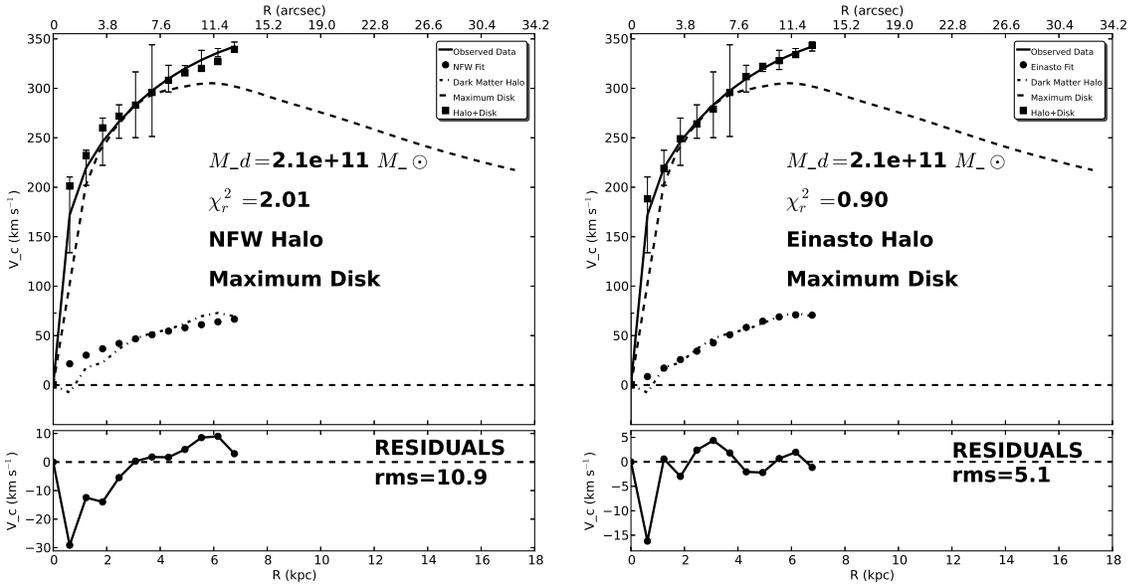}
\caption{Maximum disk solution $M_d=$ 
2.1$\times$10$^{11}$ $M_{\odot}$.\label{fig14}}
\end{figure*}

In the literature there does not exist a general definition of maximum disk. According to \citet{McGaugh2005} the maximum 
disk $M/L$ ratio should not exceed the RC data, and \citet{Gentile2008} defines the maximum disk as a fit where the stellar 
disk can totally account for the RC data peak. In the case of NGC 5278 the RC maximum is 342 km $s^{-1}$ and if we increase
the stellar disk in order to surmount the RC maximum, practically the stellar disk does not fit any point of the observed RC
of NGC 5278, so we decided to use the definition of \citet{McGaugh2005} to explore the maximum disk solution for NGC 5278, 
increasing the baryonic disk we have determined in section~\ref{sec4} by 195$\%$ to fit the majority of the internal points 
of the observed RC of NGC 5278 without overtaking the observed RC maximum. We subtracted the determined maximum disk from the 
observed RC of NGC 5278 to obtain the DM RC in the case of the maximum disk hypothesis and used the same DM halos fitting 
functions we have already used throughout this analysis to fit the DM RC. For this purpose we set the DM halos radius far from 
($\sim$ 60 kpc) the last measured point (6.77 kpc) of the observed RC of NGC 5278. The results are shown in figure~\ref{fig13}
,~\ref{fig14} and the DM halos masses are listed in table~\ref{tbl-4}. From the plots and the tabulated DM halo masses we can 
see clearly that there is still missing a certain amount of mass ($\sim$ 30$\%$) to reproduce the entire mass of NGC 5278 enclosed 
in the observed RC. Classically, the maximum disk hypothesis \citep{vanAlbada1985} have been considered to explore the possibility 
to represent the RC of some spiral galaxies only with the contribution of the baryonic component. In the case of NGC 5278 we 
already know from section~\ref{sec4} that the baryonic component cannot account for the total mass of this galaxy even in the case 
of considering more or less 30$\%$ errors in the determination of the baryonic mass. The DM analysis we have performed in 
section~\ref{sec5} tells us clearly that in the case of NGC 5278 a large amount of DM ($\approx$ 3.75 times the baryonic mass) 
component is required to entirely explain the observed RC of this galaxy and the maximum disk hypothesis confirms the conclusions 
of the RC decomposition study.

\clearpage

\section{Discussion.}\label{sec7}

At the present epoch, interactions occur in a great variety of forms and environments. In this analysis we are 
interested in first stage interactions in order to deal with the evolution of baryons and DM during the interaction 
process of NGC 5278 and its companion NGC 5279. In particular, to study the evolution of baryons and DM halos 
during an interaction process is equivalent to first analyze the effects of the DM halo physics on the galaxy disks of the 
individual pair components and then to consider the consequences induced by tidal interactions. For this reason we begin dealing 
with the repercussions of the DM distribution on the formation and evolution of isolated galaxy disks and then we examine the whole 
scenario, adding the influences of tidal forces in shaping the DM distribution in interacting galaxy pairs. The innovative work of 
\citet{Ostriker1973} demonstrates that stiff spherical DM halos decrease the self-gravity of disks and stabilize them against 
gravitational instability (i.e. bar formation). On the other hand the more recent study of \citet{Athanassoula2002} establishes 
that the formation of strong bars, due to angular momentum transfer from the stellar disk, is favored when the DM halo is 
simulated with collisionless particles and axi-symmetric. \citet{Berentzen2006} proved that triaxial DM halos create chaos in a bar
, reducing the bar instability. DM halos can efficiently slow down bars through dynamical friction \citep{Weinberg1985} and some 
authors infer the DM content of barred galaxies from the bar pattern speed \citep{Debattista1998, Debattista2000}. 
The misalignment of the baryonic infall with the minor axis of the DM halo removes a large amount of angular momentum from 
baryons during galaxy formation; as a consequence only compact spheroids can form \citep{Aumer2012}.\\ 
From the first self-consistent numerical simulations of disk/halo systems it was became clear that the DM halo play a fundamental 
role to exhaust the binding energy and angular momentum of the luminous components changing dramatically its distribution and 
dynamics \citep{Barnes1988}. Successive numerical studies of galaxy interactions attempt to constrain the DM halo mass distribution 
to the morphology of the tidal tails that generate during the dynamical evolution of the interaction process \citep{Dubinski1996}. 
\citet{Mihos1998} explore the aftermath of different DM halo potentials on the kinematics and morphology of tidal tails considering 
a dynamical model of NGC 7252 and find that halo masses of the order of 4-8 times the baryonic mass component (disk+bulge) reproduce 
the observed interaction scenario of NGC 7252; nevertheless, the same authors note that numerical simulations with cosmological 
initial conditions are needed to entirely validate the use of tidal tails to constrain the mass of DM halos in the evolution of the 
interaction process. \citet{Springel1999} perform the tidal tails test in the framework of CDM models and obtain that in the context 
of the analyzed CDM models, it is difficult, that tidal tails could constrain DM halo parameters. \citet{Springel2005a} study the 
remnants of interaction processes in the final stage of evolution at high redshift of gas-dominated galaxies by means of numerical 
simulations considering a disk and a DM halo without any bulge to determine that the rapidly cooling of the gas feeds a star-forming 
disk resembling a spiral galaxy and not an early-type object. The subject of the evolution of DM halos in interactions is too vast 
to allow a detailed treatment of the theme in this discussion, we have only briefly addressed here the most relevant points that are 
of more interest for the analysis we have made in this work. In particular we note that several results are still under debate and 
many aspects of the evolutionary problem of the DM halo in galaxy interactions are at the time totally unknown.\\ 
In this study we offer a very modest perspective that analyzes the interplay between the baryonic component of NGC 5278 and the DM 
halo of this interacting galaxy (only at the present time) through the RC decomposition technique, consisting in fitting to the 
galaxy gravitational potential different DM components, once the baryonic parameters are determined within a robust method (e.g., 
the \citet{Zibetti2009} method).  

\begin{figure*}[!htp]
\epsscale{1.0}
\plotone{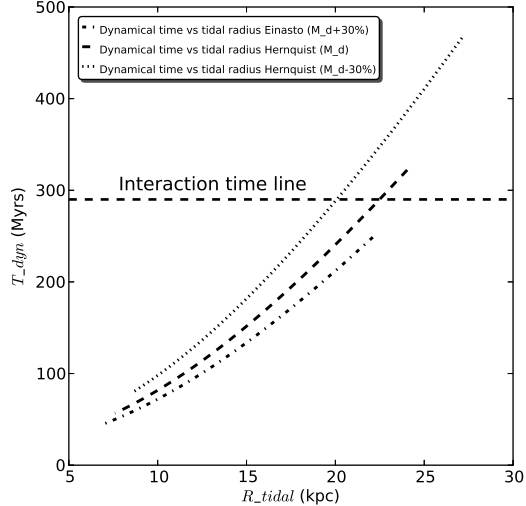}
\caption{Plot of the dynamical time 
vs tidal radius for KPG 390. The dashed 
line indicates the interaction time for KPG 390.\label{fig15}}
\end{figure*}

We perform here a simple and approximate estimate of the tidal radius and dynamical time for KPG 390 in order to assess the 
validity of our RC analysis. In our former work we have already determined the interaction time for KPG 390 to be 
$T_{int}=2.9\times10^8$ $yrs$. We consider NGC 5279 as a spherical mass distribution with total mass $m_s=2.2\times10^{10}$ 
according to \citet{Lequeux1983}, and NGC 5278 is composed by an exponential disk and a DM halo with mass $M(R)=M_d(R)+M_h(R)$, 
where $M_d(R)$ is the exponential disk mass, $M_h(R)$ represents the DM halo mass, and $R$ kpc is our RC extension. We 
consider the best fitting halos of our analysis, the Hernquist DM halo and the Einasto DM halo. The tidal radius is defined 
according to the relation $R_t=X_{12}(m_s/(k M(R)))^{1/3}$, where $X_{12}=16.6$ kpc is the separation between the two pair 
component (\citet{Karachentsev1972}), and $k=4$ indicates a circular orbit. The dynamical time is defined according to the 
relation $t_d=(2R_t^3/GM(R_t))^{1/2}$, where $R_t$ represents the tidal radius. The results are shown in figure~\ref{fig15}, and 
we can see that the minimum tidal radius begins around 6.8 kpc in the case of the cored ($n < 4$) Einasto DM halo and more 30$\%$ 
disk mass, while the observed RC of NGC 5278 extends up to 6.7 kpc; thus in principle the tidal perturbation induced by the 
companion of NGC 5278 does not affect so much the kinematics of NGC 5278 described by the H$\alpha$ RC. The dynamical time crosses 
the interaction time line beyond 20 kpc, in the case of Hernquist DM halo with disk mass $M_d=5.6\times10^{10}$ $M_{\odot}$ and 
less 30$\%$ disk mass, so that the interacting system enters the zone of interaction far beyond the extension of the H$\alpha$ RC.
From this rough guess we conclude that the RC decomposition performed in this study is not so deeply influenced by the ongoing 
interaction process between NGC 5278 and NGC 5279.\\
The observed H$\alpha$ RC of NGC 5278 reaches 6.8 kpc ($\approx$ 13 arcsecs) and the baryonic disk extends up to 17.3 kpc 
($\approx$ 33 arcsecs). The RC decomposition analysis gives us some prescriptions about which kind of DM halos better fit the 
observed RC of NGC 5278, nevertheless the short span in radius ($\approx$ 1/2.5 of the baryonic disk radius) of the RC of NGC 5278 
forces us to consider these results only as a first approximation. In fact we ignore which could be the behavior of the RC of NGC 5278
beyond the last measured point of our H$\alpha$ RC. On the other hand, it is interesting to note that in the case of disk mass 
$M_d=5.6\times10^{10}$ $M_{\odot}$, if we consider the Hernquist best fitting DM halo with final mass $M_d=8.7\times10^{11}$ 
$M_{\odot}$, we need 3.75 times the mass of the baryonic disk to account for the entire mass of NGC 5278. The necessity of such a 
large amount of DM for NGC 5278 within a radius of only 6.8 kpc is surprising given that for instance \citet{vanAlbada1985} found 
a similar DM amount (4 times the disk mass) in NGC 3198, but within 30 kpc. The same authors observe a decrease of the amount of DM 
(1.5 times the disk mass) inside 15.9 kpc. If we extrapolate that observed non linear trend to the RC of NGC 5278, we could expect 
much more DM if the observed RC could reach the baryonic disk extension, assuming a flat RC up to the optical outskirts of NGC 5278.
From these considerations we can state that even if the short range of the H$\alpha$ RC of NGC 5278 does not allow us to derive 
general properties of DM, notwithstanding it reveals the large DM content of NGC 5278 within just a few kpc.\\   
In the Introduction we have reminded that KPG 390 is classified by \citet{Klimanov2001} as M51 galaxy type. In the literature there 
are few examples of numerical modeling of these systems including the archetype M51. In particular, we are interested in trying to 
fit the coupled evolution of DM and baryons during at least the present epoch of the interaction process of M51-type interactions. 
For this reason we concisely recapitulate here the most important steps, beginning with the first attempts to reproduce this 
interacting systems, up to the actual state-of-the-art numerical experiments of this kind of interactions. The landmark 
study of \citet{Toomre1972} succeeded to reproduce the principal morphological features of M51, including bridge and tails, 
considering parabolic encounters of two galaxies idealized as disks of noninteracting tests particles without self-gravity. 
This pioneering work was unable to reproduce the grand-design spiral structure of M51 and this left out was attributed to the 
despise of self-gravity. \citet{Hernquist1990b} was the first to accomplish self-consistent 3D numerical simulations of M51 and 
focused attention on parabolic orbits. The author pointed out that the incorporation of self-gravity is not enough to generate the 
extended HI-tail discovered by \citet{Rots1990}, instead a longer duration of the perturbation is necessary to recreate the 
elongated HI feature. \citet{Toomre1994} improved the original model of M 51 adding self-gravity and increasing the duration of the 
satellite crossing since the perturbation. The result was a better morphological features matching, notably for the large HI-tail. 
\citet{Barnes1998} has noticed that the interpretation of the M51 interaction is far from being adequate, principally for the 
seemingly counter-rotating direction of the extended HI-tail, that probably denotes significant inclination with respect to the 
inner disk of M51. For this reason, according to \citet{Barnes1998} , it is complicated to produce this tilt in the framework of 
the above-quoted numerical models. \citet{Salo1993} numerically simulate the M51-type pair ARP 86 (NGC 7753-7752) considering each 
galaxy composed of self-gravitating stars+gas disks embedded in rigid spherical analytical DM halos. The resulting models match the 
morphology and kinematics of the principal tidal features and return right projected separations and radial velocity difference 
between the two pair components. A major drawback of the approach of \citet{Salo1993} is that the DM halos are treated as rigid 
components, thus the transference of material between each component of the pair cannot produce any effect on the DM distribution 
of the system, and on the other hand the DM distribution does not vary during the interaction process, fact that may affect the 
baryonic infall of one component above the other. These authors apply the same techniques in a more recent numerical study of M51 
\citep{Salo2000a, Salo2000b}. They succeeded to reproduce the outer (HI-tail) and inner morphological features of M51 and in the case 
of these studies the choice of rigid DM halos is not so crucial because the authors are not specifically interested in a detailed 
study of the relation between baryons and DM during the interaction process, yet this methodology leaves open many questions about 
the actual interplay between the main mass components in galaxies. \citet{Rosado2011} perform numerical simulations of the M51 galaxy 
type KPG 302 (NGC 3893/96), where NGC 3893 is a grand-design Sc type galaxy and NGC 3896 is cataloged as a lenticular galaxy (S0a). 
\citet{Verheijen2001} by means of HI imaging demonstrated the existence of a prolonged envelope surrounding the two galaxies of the 
pair in the direction SE to NW alongside the imaginary line connecting the nuclei of both components of KPG 302. KPG 390 is 
morphologically somewhat different from KPG 302, nevertheless we briefly summarize here the results of the numerical simulation of 
\citet{Rosado2011} to acquire more insight about the dynamical evolution of KPG 390. The numerical simulations of KPG 302 were 
achieved with GADGET2 \citep{Springel2001, Springel2005b} and using a galaxy model analogous to that of \citet{Barnes1996}. In these 
models stars and DM are treated as collisionless fluids, the gas is regarded as a compressible isothermal fluid that undergoes 
gravitational forces and hydrodynamical forces as well. 
The gas forms 10$\%$ of the stellar disk and shocks are managed including an artificial viscosity in the equations of motion. The 
gaseous and stellar disks have exponential density profiles and the DM density profiles are represented by Hernquist, P-ISO and NFW. 
The satellite galaxy NGC 3896, is represented by means of a Plummer sphere. The orbits considered for the simulations are parabolic. 
The results of the simulations resemble the observed morphology of KPG 302 and the measured H$\alpha$ velocity maps 
\citep{Fuentes-Carrera2007}. The authors outlined, it seems from the simulations that 
with P-ISO DM halos more open spiral patterns are readily formed. This fact, if confirmed with more numerical tests, probably 
indicating strong differences in shaping spiral patterns, depending on the DM density profile considered. This occurrence could be
connected with the deepness of the DM halo potential well, given that cored DM density profiles have shallower potential wells
than cuspy DM profiles; the matter in the spiral arms is rearranged at different azimuthal distances in the galaxy disk depending 
on the gravitational response of the spiral arms baryons to the collisionless DM particles. 
If this hypothesis holds at least for M51-type interactions, we would expect for NGC 5278 a less open spiral pattern due to the cuspy 
Hernquist (in the case of disk mass $M_d=5.6\times10^{10}$ $M_{\odot}$ and less 30$\%$ disk mass) DM halo gravitational potential 
well, i. e. we could constrain a range for the pitch angle of the spiral arms of this galaxy through the knowledge of the DM 
distribution. In the case of more 30$\%$ disk mass the success in the RC fitting of the cored ($n < 4$) Einasto DM halo depicts
a very different story in which probably the spiral pattern of NGC 5278 has a wider pitch angle with respect to the Hernquist DM
halo solution. At the present time these latter considerations are only on a conjectural ground and it is clear that a much more 
work is needed to deeply understand the mutual reciprocity between DM and baryons in interacting spiral galaxies.\\
In particular, in the case of KPG 390, it is difficult, at the present moment, without a numerical simulation of this pair, to 
establish an evolutionary path for the DM halo distribution in this system, due to the uncertainties about the orbital 
configuration parameters that determine the geometrical evolution of this interaction. Successive passage of the companion NGC 5279 
could diminish the baryonic content of NGC 5278 leading to very different configurations and shapes of the DM component in this 
galaxy. Summarizing, we can affirm that the present analysis does not exhaust the dynamical study of KPG 390, but settles important 
prescriptions about the DM content of NGC 5278, the primary galaxy of KPG 390. We plan to use this information as boundary 
conditions in future numerical simulations of this complex interaction process.

\section{Conclusions.}\label{sec8}

In this article we revisit the mass distribution of the spiral galaxy NGC 5278, this time with new
observational information about the disk $M/L$ coming from surface photometry and 
stellar population synthesis studies. We derive 2D $M/L$ stellar maps and find that the $M/L$ varies
along the galactocentric radius. We perform 1D and 2D bulge-disk decomposition to conclude that
our analysis does not reveal any bulge component in NGC 5278, alternatively we find out a powerful
central source and a classical exponential disk. We obtain the 2D disk mass distribution of NGC 5278 and from this 
distribution we derive the disk mass contribution to the whole RC of NGC 5278, considering a disk mass variation of 
more or less 30$\%$, to include the uncertainties of the \citet{Zibetti2009} method. We realize that the baryonic 
RC of NGC 5278, and also the baryonic RC more and less 30$\%$ disk mass, is not able to explain the whole mass content 
of NGC 5278 revealed from its gravitational potential. For this reason we subtract the baryonic disk of NGC 5278 from 
the observed RC of this galaxy, also in the cases of more and less 30$\%$ disk mass, to obtain the DM RCs for NGC 5278. 
As a next step we fit to the DM RCs of NGC 5278 four different DM halo distributions to establish which of the four DM 
halos better reproduce the DM RCs of KPG 390 A. The main results of this work are that the Hernquist DM halo better fit 
the DM RC of NGC 5278 in the case of disk mass $M_d=5.6\times10^{10}$ $M_{\odot}$ and less 30$\%$ of this value, and also 
that the cored ($n < 4$) Einasto DM halo better fit the DM RC of NGC 5278 in the case of more 30$\%$ disk mass. We also
try the maximum disk solution for NGC 5278, fitting the maximum number of internal points of the observed RC with 
195$\%$ of the determined baryonic disk. The maximum disk solution is inadequate to account for the entire mass of NGC 5278. 
We are unable to predict the time evolution of the baryonic and DM distribution because we are missing a whole dynamical study 
of the pair KPG 390. For this reason we plan to use these results on the mass distribution of NGC 5278 as boundary conditions 
in a future numerical simulation of the pair KPG 390 to unveil other important dynamic features of this complex type of 
interaction. 
 
\section*{Acknowledgments}

We would like to thank the anonymous referee for the useful suggestions and commentaries which have substantially
improved the paper. The authors wish to thank Mrs. Jana Benda for carefully reading the manuscript. This work was 
supported by DGAPA-UNAM grant IN108912 and CONACYT grant 40095-F. E. Mart\'{\i}nez-Garc\'ia acknowledges postdoctoral 
financial support from UNAM (DGAPA), M\'exico, and CONACYT's grant for postdoctoral fellowship at INAOE. This research made use 
of the NASA/IPAC Extragalactic Database (NED) which is operated by the Jet Propulsion Laboratory, California Institute of Technology, 
under contract with the National Aeronautics and Space Administration. This research has made use of the SIMBAD database, operated at 
CDS, Strasbourg, France. We acknowledge the usage of the HyperLeda database (http://leda.univ-lyon1.fr).

\end{document}